\documentclass[12pt]{iopart}

\usepackage{ltablex}
\usepackage{graphicx}
\usepackage{cite} %converts [1,2,3] to [1-3]
\usepackage{pdfpages}
\usepackage[colorinlistoftodos]{todonotes}
\begin{document}

\title[]{Case Study: Coordinating Among Multiple Semiotic Resources to Solve Complex Physics Problems}
\author{Nandana Weliweriya}
\address{Department of Physics, Kansas State University, Manhattan, KS 66506, USA}

\author{Eleanor C. Sayre}
\address{Department of Physics, Kansas State University, Manhattan, KS 66506, USA}
\ead{esayre@ksu.edu}

\author{Dean Zollman}
\address{Center for Research and Innovation in STEM Education, Kansas State University, Manhattan, KS 66506, USA} 
\address{Department of Physics, Kansas State University, Manhattan, KS 66506, USA}

\begin{abstract}
This work examines student meaning-making in undergraduate physics problem-solving. We use a social semiotic perspective to sketch a theoretical framework. The social semiotic approach focuses on all types of meaning-making practices that are accomplished through different semiotic modes that include visual, verbal (or aural), written and gestural modes and language, text, algebra, diagrams, sketches, graphs, body movements, signs, and gestures are examples for semiotic resources. We use the developed theoretical framework to investigate how semiotic resources might be combined to solve physics problems. Data for this study are drawn from an upper-division Electromagnetism I course and a student (``Larry'') who is engaged in an individual oral exam. We identify the semiotic and conceptual resources that Larry uses. We use a resource graph representation to show Larry's coordination of resources in his problem solving activity. Larry's case exemplifies coordination between multiple semiotic resources with different disciplinary affordances to build up compound representations. Our analysis of this case illustrates a novel way of thinking about what it means to solve physics problems using semiotic resources.

\end{abstract}

\maketitle

\section{Introduction} 

During problem solving, particularly at the upper division, students must coordinate multiple representations: algebraic, gestural, graphical, and verbal. While much literature in problem solving in physics focuses on introductory level problem solving, we argue that the harder problems of the upper division allow for more nuanced views of how students can connect these multiple representations to build meaning and construct compound representations which bridge multiple modes. The ability to construct representations plays an important role in helping students to make physics knowledge and communicate \cite{Prain}. The process of constructing a representation of a problem makes it easier for the problem solver to make appropriate decisions about the solution process.

In this paper, we investigate how one exemplary student solves a problem involving Ampere's Law. In the course of his problem solving, he constructs a compound representation which includes algebraic, gestural, graphical, and verbal components. To explain his problem solving, we turn to semiotic resources (Section \ref{sec:theory}). After analyzing his work in this problem (Section \ref{sec:analysis}), we discuss limitations to this approach (Section \ref{sec:discussion}) and implications for instruction (Section \ref{sec:implications}).

\section{Theory \label{sec:theory}}

%In the science education literature, `representations' refer to many ways in which students can communicate ideas, concepts, processes, and relationships. Sketches, diagrams, pictures, graphs, tables and mathematical equations are some of the commonly used representations in undergraduate physics courses \cite{rosengrant_case_2006,gire_graphical_2012,edwards_gestures_2009,radford_calculators_2003,van_heuvelen_multiple_2001,dori_virtual_2001,rosengrant_students_2009-1,scherr_gesture_2008,flood_paying_2015}. Often, representations are categorized as either internal or external representations. Internal representations refer to mental models that problem solvers need to construct within the problem-solving process \cite{rosengrant_students_2009-1}. External representations exist in the physical world and include but are not limited to sketches, diagrams, graphs, mathematical equations and words \cite{brenner_learning_1997}. 

During problem solving, students coordinate multiple representations -- algebraic, graphical, verbal, etc -- to construct arguments, abstract physical phenomena, and ultimately solve problems. Expertise in problem solving includes 
being able to represent physical phenomena using several representations and the ability to coordinate among these representations. % is one of the features of expertise in physics problem-solving and we need our students to learn and practice how to use multiple representations in their physics courses. 
Among the work has been done about students' use of multiple representations across the STEM courses, research \cite{rosengrant_case_2006, fredlund_exploring_2012-1} shows that the use of multiple representations plays a critical role in the effectiveness of the interactive engagement between the students and instructors in the teaching-learning environment. \cite{rosengrant_case_2006, flood_paying_2015, gire_graphical_2012, fredlund_exploring_2012-1}. On the other hand, the ability to translate between different representations tends to enhance students' sense-making abilities \cite{gire_graphical_2012}. 

%In addition to use of these artifacts, students engage with peers and instructors using physical objects ranging from pen and pencil to experimental apparatus and bodily actions, such as taking measurements, peer interactions and class discussions. These objects and actions play an important role in student meaning-making practices \cite{Linton, Dusen, Singh, Traver} and we interpret them as part of the representational landscape of physics problem solving. 

To understand how students make sense of and solve physics problems, we turn to social semiotic theory. Social semiotics is an approach to communication that seeks to understand how people communicate by a variety of means in particular social settings. The social semiotic approach focuses on all types of meaning-making practices that are accomplished through different semiotic modes that include visual, verbal (or aural), written and gestural modes. We couple this theory to Hammer's conceptual resources\cite{hammer2000resources} to identify small, reuseable, nameable\cite{sayre2007plasticity} chunks of student reasoning which present in a specific semiotic mode. For example, the graphical semiotic resource \textit{arrow as vector} \cite{gire2014arrows} says that students draw arrows to represent vectors, with the length of the arrow proportional to the magnitude of the vector. It combines a conceptual idea (vectors) with a particular representation (drawn arrows). 

%In the process of solving the problem, when a student uses a gesture to enact the right hand rule (connecting the directions of the inputs of the cross product with the fingers on their right hand, and making a gripping gesture to perform the cross product), we coded it as the semiotic resource (gestural) \textit{right-hand grip rule}. Similarly, we identify mathematical resources by naming them -- e.g. \textit{integral form for Ampere's law} -- for the formulae or mathematical actions they describe. Notationally, we denote resource names in \textit{italics}.

%The semiotic resource is a term used in social semiotics and other disciplines to refer to a means for meaning making. %By definition \cite{van_leeuwen_introducing_2005}, semiotic resources are ``the actions, materials, and artifacts used for communication purposes". These resources include modes such as image, writing, gesture, gaze, speech, posture; and media such as screens, 3 D forms of various kinds, books, notes, and notebooks. 

Within the context of physics problem solving, student meaning-making can be modeled as a process of using multiple semiotic resources to realize and communicate physics knowledge\cite{chapter5}. These multiple semiotic resources may come from varied semiotic modes, so the process of problem solving involves coordinating ideas across multiple semiotic modalities. The \textit{disciplinary affordance} of a given semiotic resource is ``the inherent potential of that [semiotic resource] to provide access to disciplinary knowledge.''\cite{fredlund_unpacking_2014}, allowing researchers to investigate how semiotic resources' affordances connect to disciplinary ideas. 

We are especially interested in how different affordances and constraints of different semiotic resources promote students' meaning-making as they solve physics problems. In a broad sense the research on multiple representations concern how multiple representations of scientific concepts in classroom practices affect students' understanding; but the research on social semiotic deals with students' understanding of scientific concepts through the simultaneous use of various modes of semiotic resources within and across representations.

Together with modal affordances, disciplinary affordances allow us to connect ideas in physics with the kinds of representations that best express them. As an example, spoken language is better for certain tasks and diagrams are better for other tasks; different semiotic resources access and fabricate different aspects of physics knowledge. 
%This constitutes a subtle shift in emphasis away from the particular experiences of individuals. 
Taking up disciplinary affordances allows us to focus on knowledge production and communication within the discipline (here, physics) more than focusing on the view or the experience of an individual student. 

Among the research done on the disciplinary affordances of different semiotic resources, Fredlund \cite{fredlund_unpacking_2014} uses two versions of basic RC-circuit diagrams to show the importance of unpacking the disciplinary affordance of semiotic resources for effective learning in student laboratories. First, the students were given a circuit diagram that can be connected in eight possible ways, but only one way is correct. It was difficult for students to connect the circuit and to get an appropriate output. To help students, researchers introduced a modified circuit diagram that shows the positions for connecting the signal input and the ground cable with color coded dots. The modified circuit diagram was a semiotic resource with different disciplinary affordances than the original circuit diagram. 
These new disciplinary affordances foregrounded disciplinary relevant aspects, helping students make better connections among different semiotic resources and allowing the students to make meaning of the circuit. 

In another attempt to visualize the effect of disciplinary affordances of semiotic resources, Fredlund \cite{fredlund_exploring_2012-1} investigated a group of third year physics undergraduates who are selecting among semiotic resources as they describe the refraction of light. During the task, students produced two semiotic resources: a ray diagram and a wavefront diagram. The diagrams have different potential to provide access to different aspects of disciplinary knowledge. The ray diagram could help students to reason about the refraction angles at the boundary and also about the direction of propagation, but it could not help students reason about speed changes in the two media. In contrast, the wavefront diagram promoted reasoning about speed changes but obscured reasoning about angles and directionality. 

In both of these studies \cite{fredlund_unpacking_2014, fredlund_exploring_2012-1}, Fredlund, et al. provided students with two different semiotic resources in the same semiotic mode, showing that different semiotic resources have different disciplinary affordances. In addition to the use of individual semiotic resources or selecting among two semiotic resources in student meaning making practices, research in science classrooms \cite{dori_virtual_2001,critical,fredlund_unpacking_2014, shanahan2013developing, flood_paying_2015} highlights the importance of using several semiotic resources to mediate classroom interactions \cite{Arzarello,Axelsson}. Additionally, getting students to use multiple semiotic resources helps shift their focus towards understanding the scientific processes and concepts \cite{Waldrip, kozma_roles_2000}.

\section{The Study} 

%As one of the most frequently utilized procedures in physics problem-solving, multiple semiotic resources are very useful in translating the initial, mostly verbal description of a problem into a representation more suitable for further analysis and mathematical manipulations. On the other hand, the ability to construct representations plays an important role in helping students to make physics knowledge and communicate \cite{Prain}. The process of constructing a representation of a problem makes it easier for the problem solver to make appropriate decisions about the solution process. 

\begin{figure*}[h]
\centering
\setlength\fboxsep{0pt}
\setlength\fboxrule{0pt}
\fbox{\includegraphics[width=.6\textwidth]{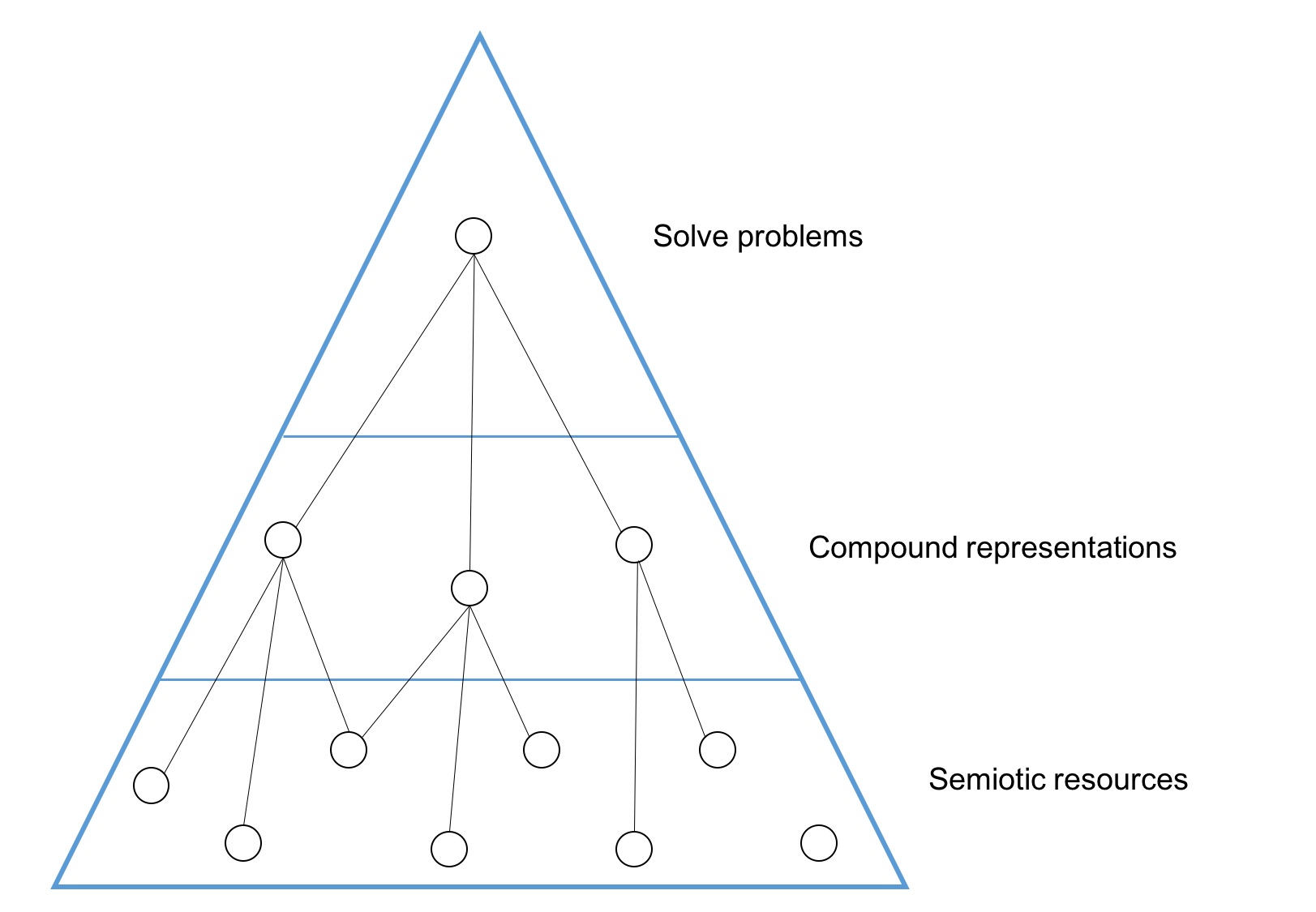}}
\caption{\label{figure1} Theoretical framework: Each semiotic resource has a semiotic mode and connects conceptual information that contributes to meaning-making. These semiotic resources are coordinated to build compound representations. Then the compound representations are used to solve the problems.}
\label{f7}
\end{figure*}

We argue that, to better represent an idea or a concept, students should be able to strategically combine multiple semiotic resources. Expanding on this idea, we adopt a social semiotic perspective to sketch a theoretical framework (Figure 1) that accounts for how semiotic resources should be combined to solve problems. In our work, we take up the idea of different semiotic resources having different disciplinary affordances. Under this framing, the process of making meaning is about coordinating different semiotic resources with different disciplinary affordances across multiple semiotic modes. By coordination we mean how the disciplinary affordances of each semiotic resource reinforce each other or hinder the use of other semiotic resources towards building representations. We introduce the idea of compound representations, composed of two or more parts or semiotic modes and linking at least two semiotic resources in the same representation. Then these compound representations can be used to solve problems. 

%however, we are interested in how students coordinate among multiple semiotic resources. 

%Students have to coordinate among different semiotic resources to understand and represent physics concepts in addition to words (speech or text). 

%Flood's \cite{flood_paying_2015} work on chemistry students shows the importance of gestures as a productive medium for representation, visualization and interaction in which students use gestures to develop their own technical language while describing molecular geometry. Chemistry experts \cite{kozma_roles_2000} use gestures, verbal language and multiple reaction diagrams to visualize the interactions among molecules to understand related chemical concepts. 

In this study, we are interested in how students coordinate among multiple semiotic resources. In particular, we consider how students \textit{construct} compound representations using semiotic resources, not how they select between researcher-provided representations. In this sense, our work extends Fredlund et al's prior work to be closer to authentic classroom practice and student problem solving. 

We explore the research question: How do students coordinate among different semiotic resources to build up compound representations while solving complex physics problems?

\section{Context and Method}

\subsection{Context}
This research was carried out at Kansas State University and the data for this study were collected from an upper-division Electromagnetism I course, which had about twenty students enrolled. As at many similar institutions, our course is a textbook-centric 4 credit-hour course with a solid foundation in the basics of theoretical physics. The class was taught by a white female instructor who had taught the same course at a different small institution. This course covers the first seven chapters of Introduction to Electrodynamics (3rd Edition) by David J. Griffiths \cite{Griffiths}. The class meets four hours per week and students work on tutorials and small group problem solving. In this course, students are highly encouraged to work in groups and think aloud while solving problems; class time is divided about equally among small group problem solving, interactive lecture, and problem-solving worksheets \cite{NguyenEJP}. Because this class has a strong focus on problem-solving, sense-making we observe instructor engages and she gives hints in a certain way that aligns with their practices in the classroom. 

As a part of the course assessments, students are required to complete two 20 to 30 minute individual oral exams with the instructor. These oral exams are used to assess students' conceptual understanding, problem-solving and scientific communication skills \cite{SayreOrals}. In this paper, we analyze the case of ``Larry" (a pseudonym), who works on an oral exam problem that takes place in the later part of the course. Larry is a strong student whose marks are near the top of his class, and his group discussions are robust and far-ranging. Larry's approach and reasoning in the oral exam is typical to a student at this level. We select Larry as an exemplary case because he is unusually verbal in oral exams compared to his peers, and we get lot of information of his problem solving activity. We picked this problem because it is a canonical problem at the upper-division level. Larry solving this problem gives us lots of insight in to how students at this level might solve this canonical problem. %This is a typical problem and Larry is an exemplary student. Here we are not looking for prevalence.
% * <dzollman@phys.ksu.edu> 2018-10-09T18:09:17.758Z:
% 
% I know the race was inserted because a reviewer asked for it. However, reviewers are not always right. In this case race is irrelevant and should not be included. Gender is obvious from the name so does not need to be included.
% 
% ^.

\subsection{Physics problem}
Larry is engaged in the problem: \textit{Suppose you had an infinite sheet which carries current $k$ equal to some constant ($k = \alpha \hat{x}$ ). What's that look like? What kind of a physical scenario is that?}. To solve this problem, one can use the right hand rule to find the direction of the magnetic field created by the sheet and to find the magnitude we could use the Ampere's law ($\oint B\cdot\mathrm{d}l =\mu_0 I_{enc}$).

\subsection{Methods}
This study has three stages of analysis. The first stage involves transcribing Larry's oral exam and dividing the problem-solving activity into segments of events. From the transcript and the video of the oral exam, each event is described without explicitly mentioning the resources (semiotic or conceptual) to generate a narrative description of the event. Within each event, we identify key elements using semiotic modes: different types of inscriptions (diagrams, mathematical formulas); extra-linguistic modes of expression (gestures); words (oral); and objects (used by Larry and the instructor). 

For the second stage of analysis, we classify semiotic and conceptual resources within each semiotic mode. First, we identify semiotic resources within inscription, verbal, and gestural modes, naming each resource descriptively. Conceptual resources are identified using published guidelines \cite{Sayre2007PhysRev}, as augmented by work in semiotic resources \cite{fredlund_unpacking_2014} and procedural resources \cite{BlackProcedural}.

Resources are named using descriptive names for the thing resources represent \cite{Sayre2007PhysRev}. For this study, we used this idea to name resources and link them with their semiotic modes.  For example, in the process of solving the problem, when the student uses a right hand rule gesture (connecting the directions of the inputs of the cross product with the fingers on their right hand, and making a gripping gesture to perform the cross product), we coded it as the semiotic resource (gestural) \textit{right-hand grip rule}. Similarly, we identify mathematical resources by naming them -- e.g. \textit{integral form for Ampere's law} -- for the formulae or mathematical actions they describe. Notationally, we denote resource names in \textit{italics}.

%As we distinguish semiotic resources, for each we give a distinct name to separate from each other (e.g., \textbf{diagrams}-\textit{parallelogram as current sheet}, \textit{line as current wire} and \textbf{gestures}-\textit{hand for wires}, \textit{pinpointing gesture}). Then to identify conceptual resources using Larry's verbal language (words), we further examine identified conceptual resources to see if there is a commonly used name in the literature. But if it does not have a commonly used name in the literature then we gave each conceptual resource a distinct name (e.g., \textit{current direction}, \textit{magnetic field direction} and \textit{current wire}). 

Finally, we make a list of conceptual and semiotic resources. The third stage of analysis involves comparing the resources' frequency and connections across episodes and connecting the semiotic resources to their disciplinary affordances (Tables 1, 2 and 3). 

We present Larry's oral exam using three episodes. We chose each episode with our research focus to show how the coordination between resources can be used to build different stages of the compound representation. We use the resource graph representation to show Larry's coordination of resources; 
%to build different stages of the compound representation 
and in our resource graphs, each circle represents either a semiotic or a conceptual resource. Within Larry's problem solving activity, we also look at the extent of his confidence using the loudness of his voice, choice of words and speed of hand gestures.
% * <dzollman@phys.ksu.edu> 2018-10-09T18:17:08.982Z:
% 
% ``Voice level" is ambiguous. It could mean either volume or frequency.
% 
% ^.

We adopt a case study methodology because it allows us to gain an understanding of the subject (Larry), events and processes that are involved in the data through a detailed analysis. Three researchers worked together in the process to establish reliability of coding and to list the disciplinary affordances of each semiotic resource that Larry uses. 

\section{Analysis \label{sec:analysis}}
 
In this section, 
%we start our analysis by identifying semiotic resources and other conceptual resources Larry uses in this problem-solving activity. Then we extend our analysis to show how Larry coordinates among semiotic resources that have different disciplinary affordances to build up a central compound representation . In the process, 
we present Larry's oral exam using two episodes, and we use the resource graph representation to show Larry's coordination of resources to build different stages of the compound representations. During the first half of his oral exam, Larry works to determine the direction of the magnetic field created by the current sheet and then continues to find the magnitude. 

\subsection{Episode 1}
The episode starts with the instructor presenting the problem. 
\begin{description}
\item[Instructor] : Suppose you had an infinite sheet which carries current $k$ equal to some constant (records on the board, $k = \alpha \hat{x}$). What's the magnetic field look like?
\end{description}

The problem statement activates the conceptual resources \textit{current sheet} and \textit{magnetic field direction}. This leads Larry to represent the current sheet on the board (Figure 2.a) using the semiotic resource \textit{parallelogram as current sheet}. After visually representing the current sheet, Larry looks at the mathematical equation ($k = \alpha \hat{x}$) on the board. The directional information ($\hat{x}$) embodied in the equation prompts Larry to think of a way to represent this detail. He uses the visual semiotic resource \textit{coordinate system} to add the directional information on to his diagram and draws three arrows with their tails together, labeling them $\hat{x}$, $\hat{y}$, and $\hat{z}$.

\begin{description}
\item[Larry] : So the coordinate \dots \textit{x} hat, \textit{y} hat, \textit{z} hat, \textit{xyz} I mean.
\end{description}

\begin{figure*}[h]
\centering
\setlength\fboxsep{0pt}
\setlength\fboxrule{0pt}
\fbox{\includegraphics[width=1\textwidth]{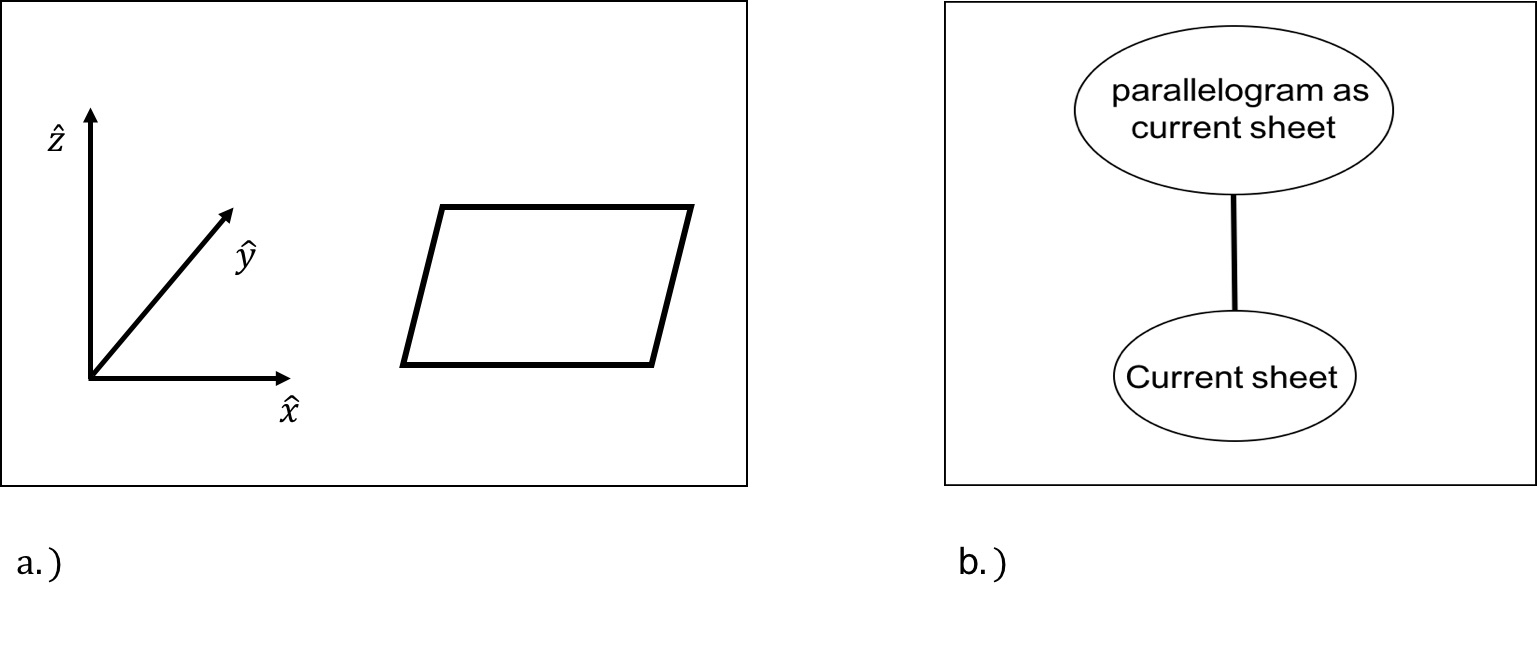}}
\caption{\label{figure2} a.) Initial compound representation b.) Resource graph for the initial compound representation.}
\label{f7}
\end{figure*}

Figure 2.a shows the initial compound representation which has a coordinate system and a parallelogram representing the current sheet. Figure 2.b shows the resource graph, the resources (\textit{current sheet} and \textit{parallelogram as current sheet}) that are connected to produce the initial compound representation. In our representation of resource graphs, each circle represents either a semiotic or a conceptual resource. After generating the initial compound representation, Larry moves to describe how he imagines the current sheet is built up from a collection of wires. Larry has the idea of focusing on a single wire to find the magnetic field that can represent the effect of the whole current sheet (\textit{part of a whole}). 

\begin{description}
\item[Larry] : um\dots{} m okay, so think of this, a sheet is kind of having a bunch of infinite wires (up and down movement of hand) one next to each other and the current from a single wire curls around this (applies right-hand grip rule).
\end{description}

Larry starts using the gestural semiotic resource \textit{hand for wires} in free space to show adjacent current wires. Then he focuses on a single wire and applies the gestural semiotic resource \textit{right-hand grip rule} to figure out and then to demonstrate the \textit{magnetic field direction} from a single \textit{current wire}. While applying the right-hand grip rule, Larry does not indicate a specific current direction (whether or not he uses the given current direction). Soon after, Larry decides to use the diagram on the board (Figure 2.a) to continue with this argument and concentrates on a point above the current sheet by using the semiotic resource \textit{pinpointing gesture}.

\begin{description}
\item[Larry] : So, I think that like if you look at a point above it (pinpoints to a location), then uh\dots{} from a single wire \dots{} . will pointing \dots{} 
\end{description}

After pinpointing to a location above the sheet, Larry considers an imaginary wire to apply the \textit{right-hand grip rule}. He applies the \textit{right-hand grip rule} for the second time without specifying the current direction and on this occasion, we observe Larry gets stuck. Instead of making a clear conclusion, he ends up repeatedly changing the orientation of his right-hand gesture.

 In order to apply the right-hand grip rule, Larry has to have a certain current direction; but when he applies the right-hand grip rule in free space, he should not have to specifically mention the current direction. Because at that point the semiotic resource \textit{hand for wires} allowed Larry to show the existence of a current wire in space, the right-hand grip rule helps him to get the \textit{magnetic field direction} using an arbitrary current direction. When Larry moves to build his argument using the diagram on the board, the diagram itself contains a coordinate system that affords to define the direction in space. So, unlike using free space, Larry has to specify the current direction before applying the right-hand rule. The missing detail of specific current direction leads Larry to decide between the orientations of his gesture. Here the semiotic resource \textit{coordinate system} hinders the use of gestural semiotic resource \textit{right-hand grip rule} without a specific current direction. 
 
 Finally, Larry decides to add the current direction information to his diagram and uses the semiotic resources \textit{arrow as vector} to represent the current direction. 

\begin{description}
\item[Larry] : So, if the current is in $x$ hat (records an arrow on diagram)
\end{description}

The coordinate system on the compound representation (Figure 2.a) allows Larry to show the direction in space (Table 1). That also permits representing the given current direction along the $x$-axis (Figure 3.a) using the semiotic resource \textit{arrow as vector} along with the mathematical symbol ($k$).

\begin{description}
\item[Larry] : Then above the sheet it would be pointing out of the board (records on diagram), from one wire, so above one wire. So, I think it would be true for the rest of the sheet as well.
\end{description}

\begin{figure*}[h]
\centering
\setlength\fboxsep{0pt}
\setlength\fboxrule{0pt}
\fbox{\includegraphics[width=1\textwidth]{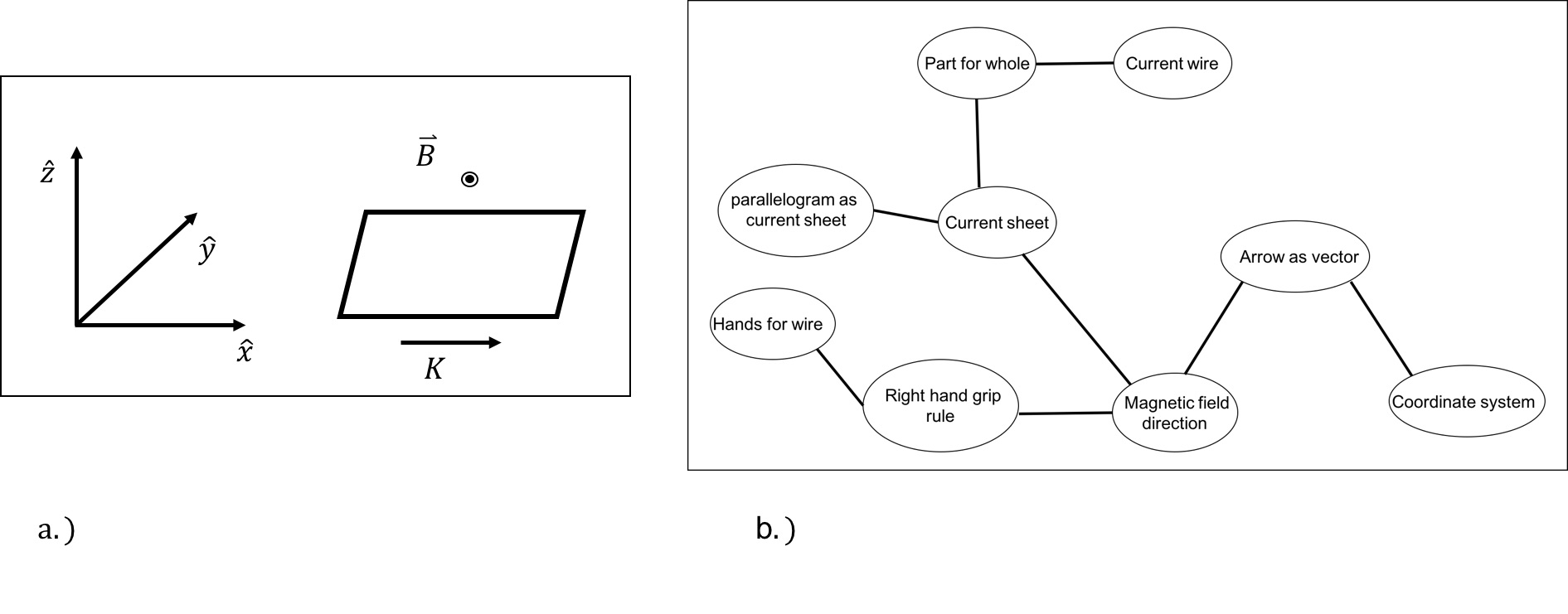}}
\caption{\label{figure3} a.) Compound representation at the end of episode 1 b.) Resource graph for this compound representation.}
\label{f7}
\end{figure*}

After recording the current direction, the rest is straightforward for Larry. He repeats the same procedure and focuses on a single wire above sheet to apply \textit{right-hand grip rule} once more. Then he uses the semiotic resource \textit{arrow as vector} to record the resulting \textit{magnetic field direction} on the diagram (Figure 3.a). Finally, Larry refers back to his initial assumption (\textit{part of a whole}) and concludes ``I think it would be true for the rest of the sheet as well.''

Figure 3.a shows the compound representation at the end of first episode, which has a coordinate system, a parallelogram to represent the current sheet, arrows to represent current direction and the net magnetic field direction above the current sheet. Figure 3.b shows the resources that are connected to produce the compound representation in Figure 3.a. We can see Larry brings in and combines more resources as he progresses with the problem at hand. At the beginning of this episode, Larry does not have a diagram to start with; but after few steps, Larry builds up a compound representation that contains a coordinate system and a parallelogram for the current sheet. Then he adds an arrow to represent the current direction with the help of the coordinate system. Larry builds up his compound representation to include more details in it. By the end of this episode, Larry adds the magnetic field direction onto his diagram by thinking of a single wire and applies the right-hand grip rule to figure out the direction. 

\subsection{Episode 2}
After finding the \textit{magnetic field direction} above the current sheet, Larry could have continued with the same argument using the diagram on board to figure out the direction below the sheet. However, the instructor introduces a sheet of paper to represent the current sheet. Then Larry switches to the sheet of paper and successfully reconstructs the magnetic field direction above sheet from a single current wire.

\begin{figure*}[h]
\centering
\setlength\fboxsep{0pt}\textbf{}
\setlength\fboxrule{0pt}
\fbox{\includegraphics[width=.5\textwidth]{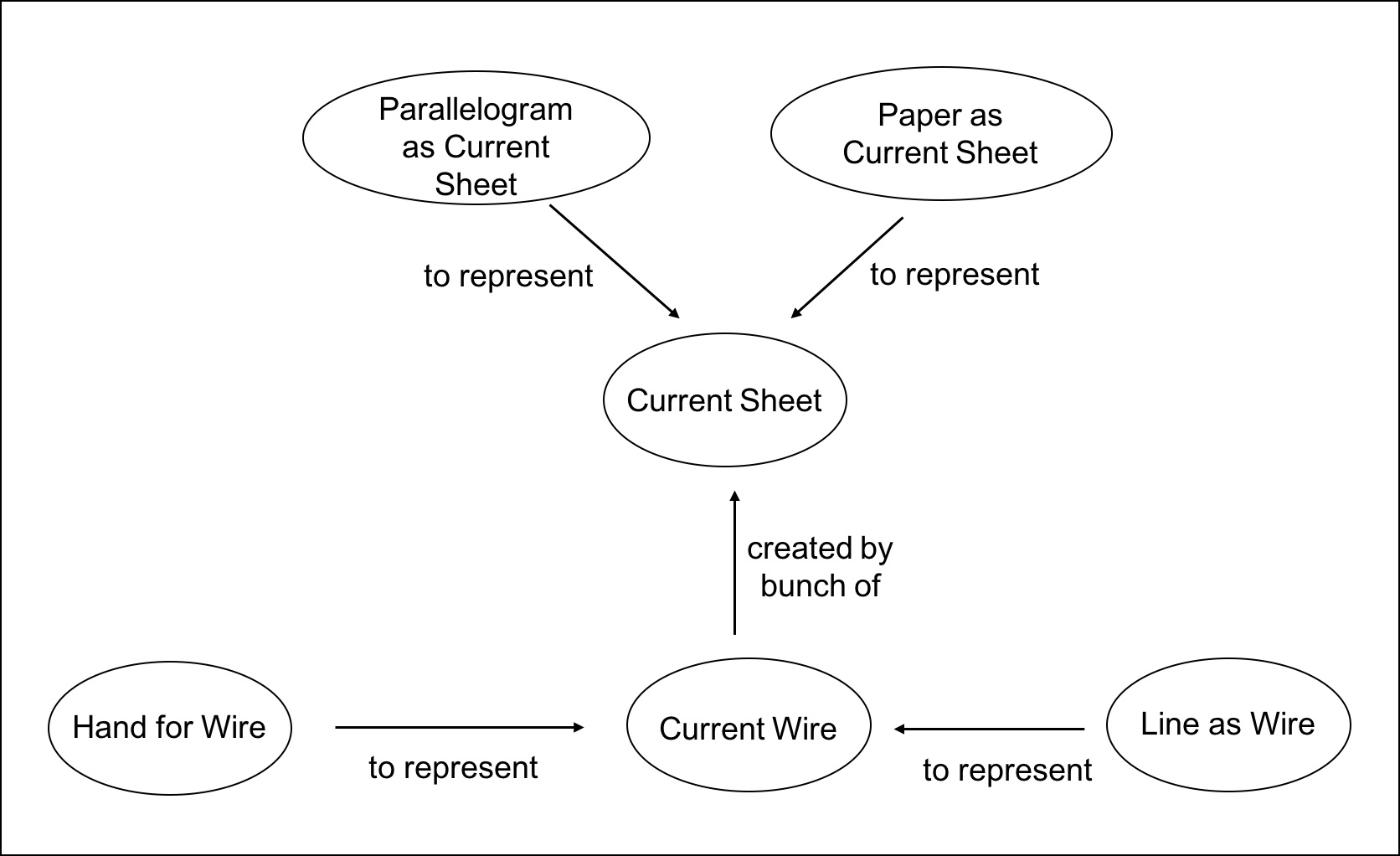}}
\caption{\label{figure2} The compound representation of the current sheet is built and represented through the gesture, sheet of paper and a drawing of a parallelogram on the board with a line to represent the current wire. }
\label{f7}
\end{figure*}

The current sheet (Figure 4) is represented using a parallelogram with a line to represent the current wire, a sheet of paper and some hand gestures. In this compound representation, the parallelogram and the sheet of paper (Table 1) allow Larry to generate visual representations of the current sheet. The inclusion of hand gestures helps to simplify the structure of the current sheet and the inclusion of the line to represent current wire further develops the idea into a visual representation. 

Table 1: Disciplinary affordances of the semiotic resources coordinated to build the compound representation of the current sheet.
\begin{tabularx}{1\textwidth}{|l|X|}
\hline
 Semiotic Resource & Disciplinary Affordance \\
 \hline
 Parallelogram as current sheet & This allows Larry to generate a visual representation of the current sheet in the space of the board. Larry keeps referring back to it as he progresses on this task, and it allows Larry to add features (current direction, current wire, Amperian loop) and findings (magnetic field direction) on to his compound representation. These recording steps also help Larry to keep track of his problem solving activities.\\
\hline
Paper as current sheet & The paper allows Larry to generate a visual representation of the current sheet in the free space. Larry refers to this while considering multiple current wires in free space and also while gesturing for the loop in episode 3. \\
\hline
Hands for wire & Larry repeats the up and down movement of his hand to show the current wires in free space. This gesture allows Larry to simplify the sheet to a bunch of wires and focus on a single current wire to apply the right-hand rule to get the magnetic field from a single current carrying wire. \\
\hline
Line as wire & This helps Larry to visually represent the current wire in the space of the sheet in the diagram (on the board) and allows him to locate the wire on the sheet. It also helps Larry to keep track of his reference current wires (with embodied current direction) while applying the right-hand grip rule, which leads Larry's effort to a successful conclusion. \\
\hline

\end{tabularx}

\begin{description}
\item[Larry] : So, if this is the sheet (refers to the sheet of paper), the current is going this way (across the surface of sheet- right to left), and looking at a point above it, then from one wire, magnetic field will be point in that way (away from him). 
\end{description}

Larry first uses the semiotic resource \textit{paper as current sheet} to represent the current sheet in free space. Then he uses the gestural semiotic resource \textit{finger pointing in direction} to specify the current direction and directs his index finger along the current direction. Finally, he applies the \textit{right-hand grip rule} by considering an imaginary current wire that is located above the sheet. While using the sheet of paper to figure out the \textit{magnetic field direction} above the sheet, we can observe Larry's confidence as he reasons smoothly and his voice stays around the same volume. This may be because this step is all about repeating the same procedure, even though now he is using a different semiotic resource.

Next, Larry decides to take his argument to the next level and starts to reason for the net magnetic field direction above the current sheet. Larry moves to consider multiple current wires and he starts by locating two imaginary current wires using the semiotic resource \textit{pinpointing gesture}: one closer to the surface of sheet and the other little above the surface of sheet. Then he applies the \textit{right-hand grip rule} by considering one wire at a time, but quickly gets stuck.

\begin{description}
\item[Larry] : So like from this wire (above the surface of sheet), the current is gonna \dots{} the magnetic field points that way (away from him), but from a wire over here (closer to the surface of sheet), it would be pointing more \dots{} uh\dots{} this way (pointing towards him), uh\dots{} so it seems like they are gonna superposition, kind of complicated.
\end{description}

After applying the \textit{right-hand grip rule} for the second wire, Larry realizes that the resulting magnetic field directions are in opposite directions, which contradicts his previous conclusion (``I think it would be true for the rest of the sheet as well''). Previously he concluded that above the current sheet, the \textit{magnetic field direction} from all the individual current wires should be in the same direction. Larry's explanation about the net effect reaches a dead end (``uh \dots{} so it seems like it's gonna superposition, kind of, complicated.'') At this point, Larry's voice volume indicates a confusion. We can observe Larry's strong voice starts to fade after considering the second wire, his pauses between words increase and the speed of his hand movements decreases. In addition, his words ``it would be \textbf{ pointing more} \dots{} uh \dots{} this way \dots{} \textbf{kind of complicated}'' show his uncertainty about the result. 

As a result of this confusion, Larry returns to the diagram on the board to consider multiple current wires. First, he uses the semiotic resource \textit{line as wire} (Figure 6.a) to add a current wire on to his diagram. 

\begin{figure*}[h]
\centering
\setlength\fboxsep{0pt}\textbf{}
\setlength\fboxrule{0pt}
\fbox{\includegraphics[width=.6\textwidth]{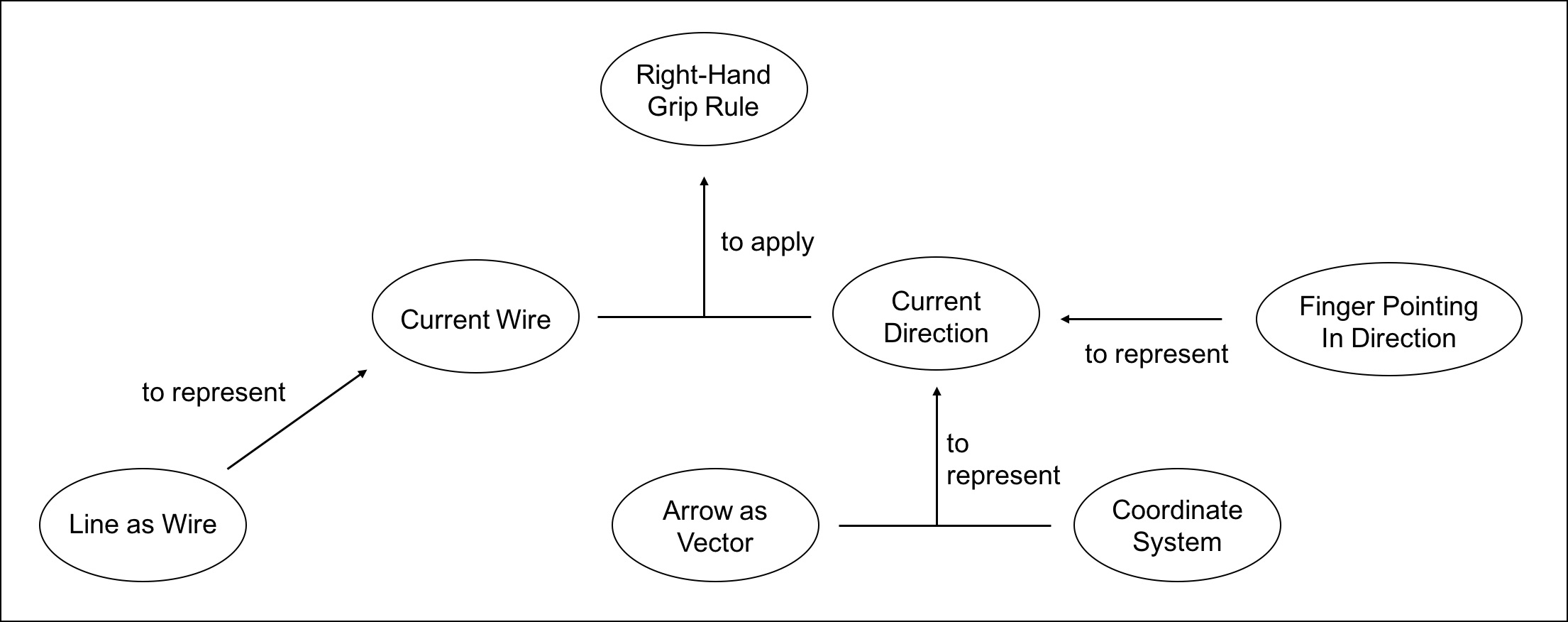}}
\caption{\label{figure5} The compound representation of the right-hand grip rule is developed using gesture and drawing of a line to represent the current wire with an arrow to represent given current direction and is applied using the right-hand gesture.}
\label{f7}
\end{figure*}

The compound representation of right-hand grip rule (Figure 5) allows him to figure out the resulting direction of the magnetic field lines from a current carrying wire. On some occasions, a gesture (table 2) allows Larry to specify the current direction before applying the right-hand grip rule. Later, the coordinate system (Table 2) allows Larry to represent the given current direction (using an arrow), and the line to represent current wire allows Larry to apply right-hand grip rule to find the magnetic field direction.

Table 2: Disciplinary affordances of the semiotic resources coordinated to build the compound representation of the right-hand grip rule.
\begin{tabularx}{1\textwidth}{|l|X|}
\hline
 Semiotic Resource & Disciplinary Affordance \\
 \hline
Coordinate system & A coordinate system enables Larry to define the location of a point, distance between points, and direction in space (on a planar surface). This allows Larry to represent the given current direction (using an arrow) that he uses to find the magnetic field direction.\\
\hline
Arrow as vector & Arrow as vector permits Larry to visually represent the vector direction. As the coordinate system permits showing the direction in space, using an arrow to represent a vector allows Larry to represent the given current direction along the $x$-axis. In this situation, visually representing the current direction helps Larry to apply the right-hand rule towards determining the magnetic field direction. Later, the same semiotic resource allows Larry to visualize the resulting magnetic field directions. \\
\hline
Finger pointing in direction & Larry uses the index finger to the side while gesturing for direction. This allows him to show the direction of the current on the surface of paper and also in the space of the sheet in the diagram (on the board) that helps to apply right-hand grip rule. \\
\hline
\end{tabularx}

\begin{description}
\item[Larry] : If we are looking at a wire right here (draws a line at middle of the sheet), and then, so the current above the wire (applies right-hand grip rule) my hand curls, points back in me. Uh \dots{} if we look at a wire, like this is an infinite sheet so way back here in the y direction, uh then my fingers at that point are more pointing in down than they are towards me.
\end{description}

\begin{figure*}[b]
\centering
\setlength\fboxsep{0pt}
\setlength\fboxrule{0pt}
\fbox{\includegraphics[width=.95\textwidth]{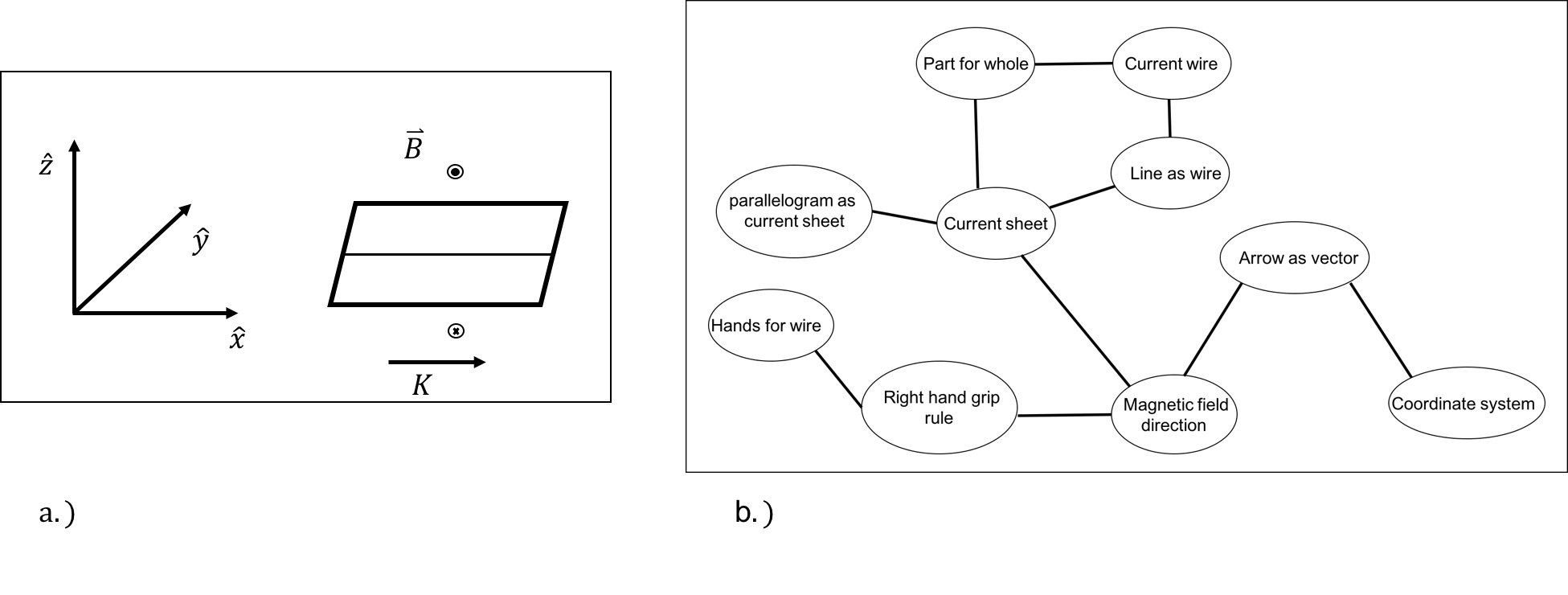}}
\caption{\label{figure6} a.) Compound representation at the end of episode 2 b.) Resource graph for this compound representation.}
\label{f7}
\end{figure*}

Larry considers the current-carrying wire that he just recorded on the diagram to apply the \textit{right-hand grip rule} and successfully reproduces the same result. Then, Larry considers a second current wire located far back in $y$ direction and applies the \textit{right-hand grip rule}. Switching to the diagram on the board and recording a line to represent the current wire allows Larry to consider two current wires when drawing a conclusion about the net \textit{magnetic field direction} above sheet. The semiotic resource \textit{line as wire} helps Larry to explicitly indicate the location of the current-carrying wire on the diagram. In addition, this inclusion allows Larry to incorporate and specify the current direction (which is already recorded on diagram) while applying the right-hand grip rule. When Larry switches to the diagram on the board, the semiotic resources \textit{line as wire} and \textit{arrow as vector} allow Larry to keep track of his reference current wires along with the current direction. But while using the sheet of paper to consider multiple wires, Larry works in free space using the gestural semiotic resources \textit{pinpointing gesture} and \textit{finger pointing in direction} to locate imaginary current wires and to show the current direction. While working in free space, in addition to keep tracking of the wire he is considering, Larry has to remember more than one direction at a time (current and resulting magnetic fields), this leads Larry's effort to a conflict. 

The instructor's follow-up question leads Larry to consider another wire located far forward in y direction (\textit{mirror current wire}) to make a conclusion about the net effect above the sheet. 

\begin{description}
\item[Instructor] : What about one, that mirror, so that wire way far back, the mirror wire way far forwards? 
\end{description}

After reapplying the \textit{right-hand grip rule}, Larry reasons for the net direction above the current sheet He says ``Uh, my finger is pointing up''. This result aligns with his earlier conclusion. Then the instructor suggests that Larry consider below the sheet as well, asking ``So, it came out above and below it goes?''
Larry continues with the same argument and reapplies the \textit{right-hand grip rule}. Then he concludes that ``so below they go in" and uses the semiotic resource \textit{arrow as vector} to record the magnetic field direction below the sheet on diagram (Figure 6.a).

At the beginning of episode 2, the compound representation (Figure 3.a) has a coordinate system, a parallelogram to represent the current sheet, arrows representing the magnetic field direction above sheet and the current direction. Now by the end of episode 2, Figure 6.a has all the required information about \textit{magnetic field direction} above and below the current sheet. Larry has added a line to represent a current-carrying wire and an arrow to represent net magnetic field direction below the current sheet. Figure 6.b shows the resources that are connected to produce this compound representation and we see Larry continues with the same combination of resources as he is still looking for the magnetic field direction. As the only modification, he adds the \textit{line as wire} to represent the current wire and further builds up his compound representation to include magnetic field direction below the current sheet. 

\subsection{Episode 3}
Larry completes the first task by finding the magnetic field direction and now he has to figure out the magnitude of the magnetic field. To start the process, Larry records the semiotic resource of mathematical formula for the \textit{integral form of the Ampere's law} ($\oint B\cdot\mathrm{d}l =\mu_0 I_{enc}$). This mathematical formula visualizes the relationships between the integrated magnetic field around a closed loop and the electric current passing through the loop. In order to continue with the mathematical manipulations, Larry has to pick an Amperian loop, pick the dimensions of his loop and then pick a direction for his loop. Instead, Larry does this as a two-step process. First, he gestures for the loop orientation and tries to advance with mathematical manipulations. Later as he gets stuck, he picks the dimensions and a direction for his loop.

After recording the mathematical formula for the Ampere's law, the instructor suggests Larry pick a loop,

\begin{description}
\item[Instructor] : Cool, now you need to pick an Amperian loop.
\item[Larry] : I'll pick a loop. Current is in this way (pointing on the surface of paper). I think the loop is like this (hand shows the loop perpendicular to the edge of the paper).
\end{description}

Within this step Larry combines several semiotic resources. First, he returns to the semiotic resource \textit{paper as current sheet} and uses it along with the gestural semiotic resources \textit{finger pointing in direction} and \textit{hand for loop} to show how he picks the loop. The semiotic resource \textit{paper as current sheet} allows him to represent the existence of the current sheet in free space, \textit{finger pointing in direction} helps to indicate the current direction. Then, the semiotic resource \textit{hand for loop} helps to visualize the orientation of the loop with respect to the sheet of current. But, it does not help Larry figure out the vector orientation between the magnetic field and the unit length on the Amperian loop that is required to manipulate the integral in the Ampere's law equation. Later, we observe this limitation of the semiotic resource \textit{hand for loop} leads Larry to get stuck, and then Larry uses another semiotic resource to represent the orientation of loop.

Larry's intention is to continue with the mathematical manipulation, so after gesturing for the loop, he moves to figure out the \textit{current enclosed} (current flowing through the Amperian loop).

\begin{description}
\item[Larry] : My current enclosed is gonna be \dots{} said got (records equation) ($k = \alpha \hat{x}$), so my current is just, my current enclosed is just gonna be uh \dots{} $k$ $dl$ right (records $I_{enc}=k\dots{} $). But it can't just be $k$ times $dl$ cause that's a vector, and it not be a vector. I mean two end up is not a vector. But, could it be $k$ dot $dl$, you like the sound of that? 
\end{description}

Although he is ultimately unsuccessful, Larry expresses certainty that he could use a mathematical formula to get the current flowing through the loop. So far, Larry has not picked dimensions for his Amperian loop, instead he just uses the hand to show the orientation of the loop. This limitation prevents his moving forward with the mathematical manipulation. As Larry gets stuck, we observe Larry's voice keeps fading and he pauses between words more than usual. Finally, he asks for the instructor's feedback to move forward: ``could it be \dots{} . you like the sound of that?''. Instead of answering Larry's question, the instructor suggests he add the loop to his diagram, asking him ``So how big is your loop? Draw your loop.'' After the instructor's suggestion, Larry uses the semiotic resource \textit{square as loop} to add a loop on to his diagram (Figure 8.a).

\begin{figure*}[h]
\centering
\setlength\fboxsep{0pt}\textbf{}
\setlength\fboxrule{0pt}
\fbox{\includegraphics[width=.5\textwidth]{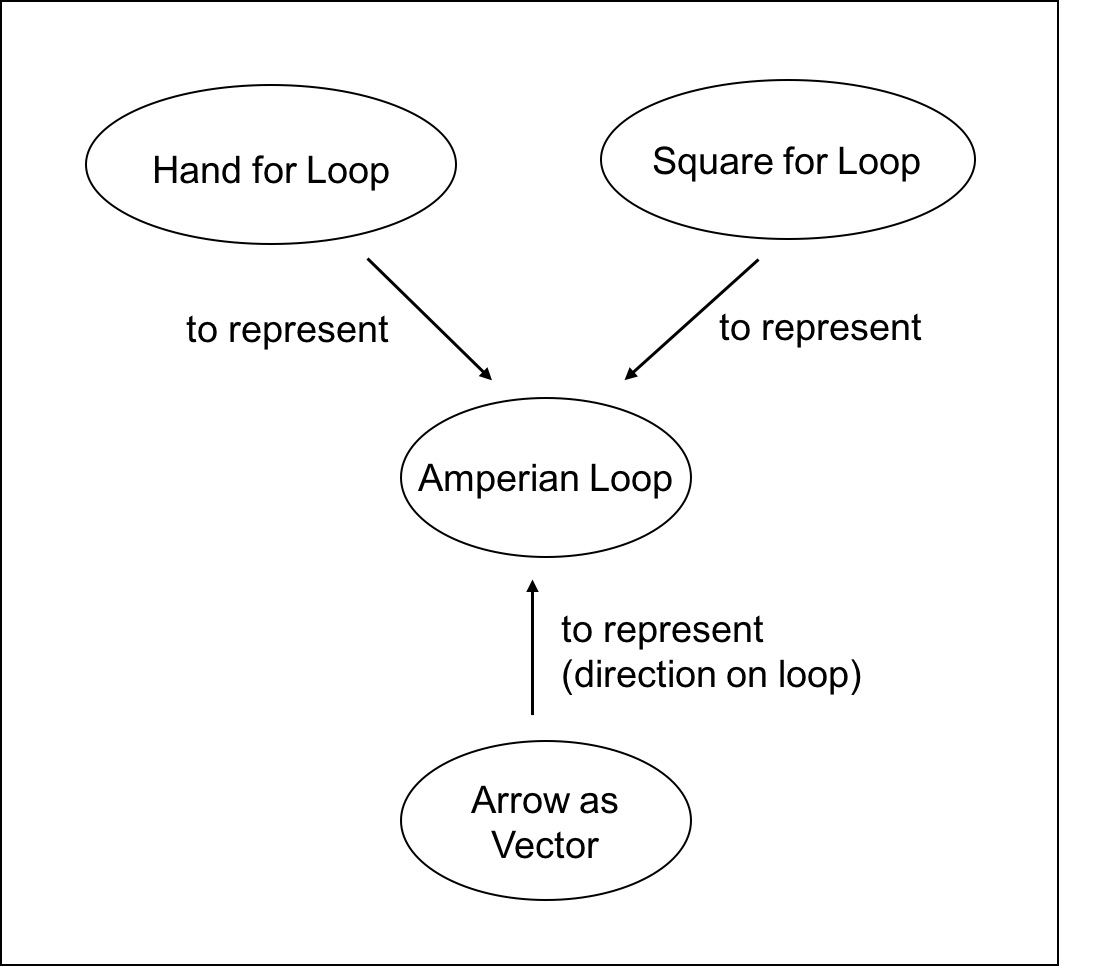}}
\caption{\label{figure7} The compound representation of the Amperian loop is built and represented through the gesture and a drawing of a square on the board with an arrow to represent the direction on the loop.}
\label{f7}
\end{figure*}

In the compound representation of the Amperian loop (Figure 7) gesturing and drawing (square) on the board (Table 3) allows Larry to visually represent the existence of the loop with respect to the orientation of the current sheet. Then the inclusion of the direction on the loop helps to figure out the relative orientation with the magnetic field direction.

Table 3: Disciplinary affordances of the semiotic resources coordinated to build the compound representation of the Amperian loop.
\begin{tabularx}{1\textwidth}{|l|X|}
\hline
Semiotic Resource & Disciplinary Affordance \\
\hline
Hand for loop & Larry uses the hand to represent the location and orientation of the loop with respect to the orientation of the paper sheet. Even after recording the loop on the diagram, Larry reuses the hand for loop to best communicate his idea to the instructor. \\
\hline
Square for loop & This helps to generate a visual representation of the loop in the space of the board (diagram on the board) and allows Larry to show the orientation of the loop with respect to the orientation of the current sheet (parallelogram for sheet). This recording step helps Larry to continue with the mathematical manipulations as he labels the loop dimensions and to figure out the current enclosed in the Amperian loop. \\
\hline
Arrow as vector & Arrow as vector permits Larry to visually represent the given current direction and the resulting magnetic field directions. While Larry is trying to simplify the left-hand side of the integral form of the Ampere's law equation during the latter part of episode 3, either gesturing for the loop or just recording the loop does not help Larry to figure out the relative orientation between the magnetic field direction and the unit length on the Amperian loop. Then the use of arrow to represent the direction on the loop helps Larry to build a complete argument and advance with the mathematical manipulation.\\
\hline
\end{tabularx}
% * <dzollman@phys.ksu.edu> 2018-10-09T18:51:44.228Z:
% 
% I think you mean Table 3 in the paragraph above the table. The Table 3 caption does not show up on the PDF copy. I cannot see why.
% 
% ^.

Before moving forward, Larry decides to explicate the orientation of the loop because he doubts his drawing ability. To gesture the orientation of the loop, he uses the semiotic resources \textit{paper as current sheet}, \textit{finger pointing in direction} and \textit{hand for loop}. Next, the instructor reminds Larry to pick the dimensions for his loop.
% * <dzollman@phys.ksu.edu> 2018-10-09T18:57:13.788Z:
% 
% The following down to ``Larry: So \dots{} " does not appear in the PDF
% 
% ^.

\begin{description}
\item[Instructor] : Okay, good. And how wide is your loop? And what is the other dimension?
\end{description}

Larry labels the loop dimensions as ``$l$" and ``$w$" (Figure 5.a), and this step allows Larry to figure out the current flowing through the loop that he could not do earlier. He records the semiotic resource of mathematical formula for \textit{current enclosed}, $I_{enc}=kl$ and substitutes the given current information ($k = \alpha \hat{x}$) for current density $I_{enc}=kl=l\alpha \hat{x}$. Even though Larry correctly reasons that ``it ($I_{enc}$) not be a vector. I mean two end up is not a vector'', his mathematical formula for current through the loop still has the vector information ($\hat{x}$). Later, Larry is going to realize that he does not need to have $\hat{x}$ anymore because he does the dot product.

After figuring out the current through the loop (\textit{current enclosed}), Larry moves towards simplifying the left-hand side of the Ampere's law formula. First, he records the semiotic resource of mathematical formula $\oint B\cdot\mathrm{d}l =\mu_0 \int k\cdot\mathrm{d}a$. 

\begin{description}
\item[Larry] : So (records on the board) integral $B$ dot $dl$ equals $\mu_0$ integral $k$ dot $da$. And then \dots{} for this (left hand side) I can do four separated integrals right? Since its square (show by hand) and these two (``$w$" legs on loop) do not end up mattering. 
\end{description}

Larry calls out loud the names of the mathematical symbols while recording the formula and then uses the gestural semiotic resource \textit{hand for loop}, while talking about the left-hand side of the Ampere's law equation. Larry gestures that the loop has four sides and then uses the semiotic resource \textit{pinpointing gesture} to indicate the two ``$w$" legs (Figure 8.a). Without providing complete reasoning, Larry talks about how the integral manipulation regarding two ``$w$" legs cancels out. Then the instructor helps Larry to build up his argument. 

\begin{description}
\item[Instructor] : Because?
\item[Larry] : Because they're perpendicular to the \dots{} Am I \dots{} Is it these ones that I am not gonna concern about? I am pretty sure it is.
\item[Instructor] : Yes. Because the direction of $dl$ and the direction of $B$ are \dots{} 
\item[Larry] :\dots{} are perpendicular. 
\end{description}

The above conversation between the instructor and Larry shows the argument behind the integral manipulation. Because the orientations of the ``$w$" legs are perpendicular to the magnetic field direction, when Larry takes the dot product they cancel out. Even though Larry's argument is correct, so far he has not picked a direction for his loop. Just the existence of loop without direction on it could not allow Larry to see the relative vector orientations and this prevents his making a complete argument. Even though Larry seems to be building an argument just by completing the missing bit of the instructor's statement, we can observe that Larry's argument is not yet complete as he moves to consider the other two legs (``$l$" legs).

\begin{description}
\item[Instructor] : Okay. What's about the other two legs? 
\item[Larry] : Uh \dots{} So for the other two legs, uh \dots{} all right, so this one is kind of above it in the z, which means that the field is coming out at me (finger pointing in direction gesture). \dots{} So, I need to pick a direction for my loop, don't I? 
\end{description}

Instead of continuing with the same argument, Larry uses the semiotic resource \textit{finger pointing in direction} to reason about the orientation of the loop in three-dimension (3D). Larry stops without making a complete reasoning. Larry then immediately, without prompting from the instructor, realizes that he needs to pick a direction for his loop (\textit{``so I need to pick a direction for my loop, don't I?"}) and uses the semiotic resource \textit{arrow as vector} to record the loop direction (Figure 5.a).

\begin{description}
\item[Larry] : uh \dots{} Let's say it all goes this way (counter clockwise). Okay. So, above it $B$ is coming out and the way I draw my loop is coming out at me so they are parallel and then below it B is going in and my loop is going that way so they're parallel again. 
\item[Instructor] : Cool. OK. So, you get for that integral? 
\item[Larry] : Uh\dots{} so $2Bl$ (records on board) 
\end{description}

Either picking the loop using the semiotic resource \textit{hand for loop}, or recording the loop using semiotic resource \textit{square for loop}, does not help Larry to figure out the relative orientation between the magnetic field and the unit length on the Amperian loop. This prevents Larry from advancing with the mathematical manipulations. But the inclusion of \textit{arrow as vector} to represent the direction on the loop helps Larry to continue with a valid argument. Larry reasons how the \textit{unit length} (on ``$l$" legs) on the loop is parallel to the magnetic field direction. Also, we can see Larry's confidence level as he reasons with a strong voice but with no pauses between words.

After few communication steps with the instructor, Larry recalls the mathematical formula for the current through the loop and finds the magnitude of the magnetic field to finish the task. 

\begin{description}
\item[Instructor] : How much current pierces this loop? You worked this out not two lines ago. 
\item[Larry] : Right. $l\alpha \hat{x}$ 
\item[Instructor] : Yup. Though you do not need the $x$ hat anymore because you have done the dot product. 
\item[Larry] : All right. (Completes the right side by $\mu_0 l \alpha$ and records on the board: $B =\mu_0 \alpha /2$). 
\end{description}

\begin{figure*}[h]
\centering
\setlength\fboxsep{0pt}
\setlength\fboxrule{0pt}
\fbox{\includegraphics[width=.95\textwidth]{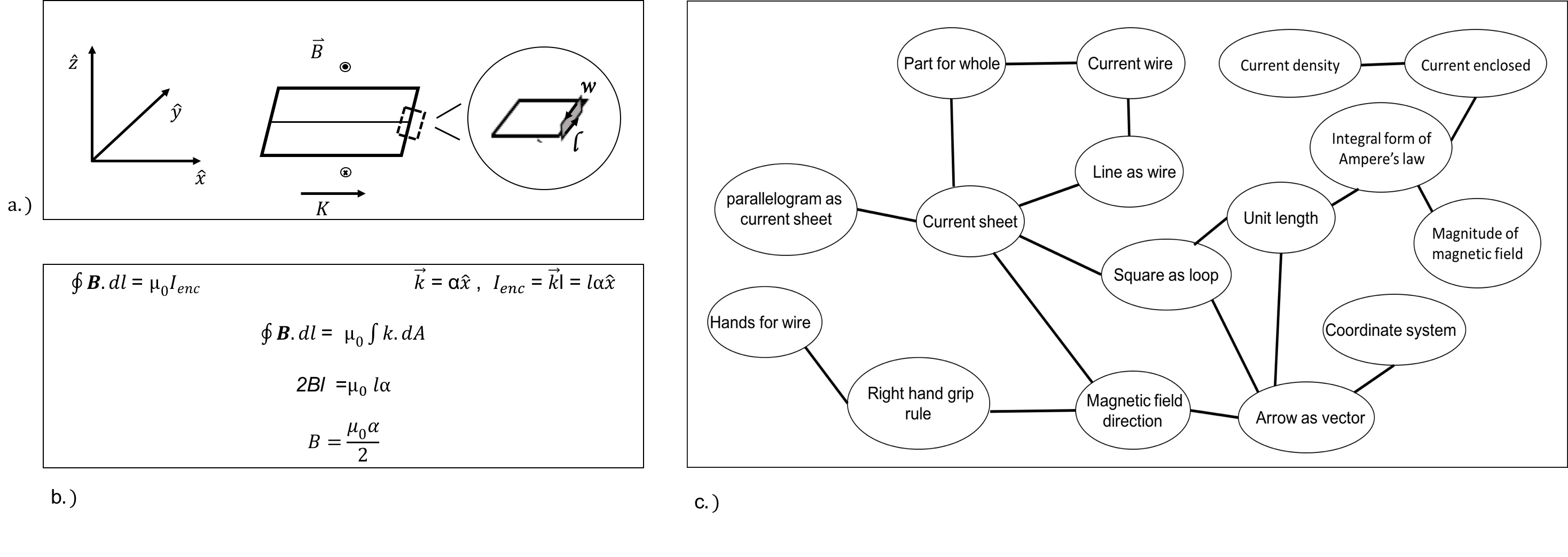}}
\caption{\label{figure8} a.) Compound representation at the end of episode 3 b.) Larry's use of mathematical formulas c.) Resource graph for the final compound representation and mathematical formulas.}
\label{f7}
\end{figure*}

Finding the \textit{magnitude of the magnetic field} is a new task for Larry. At the beginning of episode 3 Larry has a compound representation that contains a coordinate system, arrows to represent current and the magnetic field directions, a line to represent a current wire and a parallelogram to represent the current sheet. Larry starts by recording the mathematical formula for the \textit{integral form of the Ampere's law} and combines it with the other mathematical relations to get the current through the loop (\textit{current enclosed}). In order to manipulate the integral on left hand side of the \textit{integral form of the Ampere's law} formula, Larry starts by recording an Amperian loop on his digram and later he adds the details of Amperian loop dimensions and direction to his compound representation (Figure 8.a). This recording step helps Larry to manipulate the integral and find the magnitude of magnetic field (Figure 8.b). In the resource graph (Figure 8.c), we see Larry coordinate between conceptual resource \textit{unit length} and semiotic resource \textit{square as loop} along with additional resources related to mathematical formulas: \textit{integral form of the Ampere's law}, \textit{current enclosed}, \textit{current density} and \textit{magnitude of the magnetic field}. 

\subsection{Connections among the episodes}

We see the disciplinary affordances of some semiotic resources that Larry uses to reinforce the use of forthcoming semiotic resources. Larry starts his oral exam by combining semiotic resources \textit{coordinate system} and \textit{parallelogram as sheet} to build up his compound representation. The \textit{coordinate system} allows Larry to represent the current direction using \textit{arrow as vector}. This inclusion allows Larry to focus on a single current-carrying wire to apply the \textit{right-hand grip rule} to get the magnetic field direction. This finding leads Larry to further develop his compound representation by using \textit{arrow as vector} to include the \textit{magnetic field direction} above the sheet. Moving on to consider multiple current-carrying wires using \textit{paper as current sheet} results in Larry switching to his diagram on the board to use \textit{line as wire} to represent the current-carrying wire. The inclusion of \textit{line as wire} and \textit{arrow as vector} (the current direction information which is already available in the diagram) allows Larry to consider multiple current-carrying wires to make a conclusion about magnetic field direction above and below the sheet. At the end of episode 2, Larry uses \textit{arrow as vector} to further build up his compound representation to include the magnetic field direction. After determining the magnetic field direction, Larry moves to find the magnitude of the magnetic field in episode 3. Larry starts with the mathematical formula \textit{integral form of the Ampere's law} and combines it with the given information of current density to get the current through the loop (\textit{current enclosed}) to simplify the right-hand side of the integral. After manipulating the left-hand side of the Ampere's law integral, Larry further adds the details of the Amperian loop to his compound representation. He uses \textit{square as loop} and \textit{arrow as vector} to represent and to pick a direction for the loop. This recording step helps Larry to manipulate the left-hand side of the integral and finally to obtain the magnitude of the magnetic field. 

Our study also shows the disciplinary affordances of some semiotic resources hindering the use of other semiotic resources. In episode 1, Larry first applies the \textit{right-hand grip rule} in free space, without specifying the current direction to figure out the magnetic field direction. Then he continues to apply the \textit{right-hand grip rule} considering diagram on the board again without specifying the current direction. The directional information embodied on the diagram on the board by \textit{coordinate system} obstructs the use of \textit{right-hand grip rule} and Larry ends up changing his gestural orientation. Later the inclusion of the current direction information helps Larry to figure out the magnetic field direction. 

During the first two episodes, Larry works to get the magnetic field direction, and we observe him using the semiotic resource \textit{right-hand grip rule}. Once he gets the magnetic field direction and builds onto his compound representation, he no longer uses the \textit{right-hand grip rule}. As Larry moves to use the mathematical formulas in episode 3 to get the magnitude, the disciplinary affordance of mathematical formulas \textit{integral form of the Ampere's law}, \textit{current enclosed}, \textit{current density} and \textit{magnitude of the magnetic field} prevents the use of semiotic resource \textit{right-hand grip rule}. It was important but no longer useful for Larry to get the magnitude of magnetic field in the episode 3. 

Likewise, Larry uses the semiotic resource \textit{hand for loop} to gesture how he picks the Amperian loop in the third episode. Using \textit{hand for loop} helped Larry to show the orientation of the Amperian loop relative to the orientation of the current sheet. As Larry switches to work on mathematical formulas to get the magnitude of magnetic field, he ends up associating mathematical formulas: \textit{integral form of the Ampere's law}, \textit{current enclosed} and \textit{current density}. The inclusion of mathematical formulas constrains the use of semiotic resource \textit{hand for loop}. We observe Larry switching to the diagram on the board to represent the loop using a \textit{square as loop}. Just the hand gesture is not helpful in determining the current piercing the loop or in determining the left-hand side of the Ampere's law integral. 

\section{Discussion \label{sec:discussion}} 

%Student learning can be described as individuals making connections between formal ideas and everyday experiences. 
The process of student meaning-making is not a purely cognitive one; in physics classes, both students and teachers use a number of semiotic resources in addition to speech and writing. The approach \cite{chapter5} to consider the involvement of all artifacts, objects, and actions to describe student meaning-making broadens the boundaries in physics as a discipline. The idea of disciplinary affordance allows us to connect ideas in physics with the kinds of representations which best express them. 

Fredlund \cite{fredlund_unpacking_2014, fredlund_exploring_2012-1} used the affordances approach to show that different semiotic resources have different disciplinary affordances. They interviewed students on particular problems which had two specialized visual representations, each of which has a strong set of disciplinary affordances. Previous work on disciplinary affordances did not investigate how students could combine multiple semiotic resources within classroom problem-solving. In our study, we have actual classroom data, and we picked a more typical problem. As research programs, both studies so far consist of interviews with students solving very particular, but typical, problems. So, it is possible that further research could involve interviews or classroom observations with a series of typical problems to further explore how the representations are developed, determined to be insufficient and replaced or augmented by new ones brought in by the students.
% * <dzollman@phys.ksu.edu> 2018-10-09T19:21:14.194Z:
% 
% Please check the sentence, `` As research programs, both studies so far consist of interviews with students solving very particular, but typical, problems." I was not certain what you wanted to say. My editing is my best guess at your intent. 
% 
% 
% ^.

The problem Larry solves is a canonical problem at the upper-division level. Larry solving this problem gives us lots of insight into how students at this level might solve this problem. But, we are not looking for prevalence, and we are not trying to make a normative argument that all the students should solve this problem in the same way. This particular problem required Larry to start from a diagram and move to the mathematics at the later part of his solution; this process required Larry to coordinate between a multitude of semiotic resources. We observe Larry combine a series of semiotic resources with other available conceptual resources, some of which he keeps coming back to and some of which he discards after using. 

\begin{figure*}[h]
\centering
\setlength\fboxsep{0pt}\textbf{}
\setlength\fboxrule{0pt}
\fbox{\includegraphics[width=.4\textwidth]{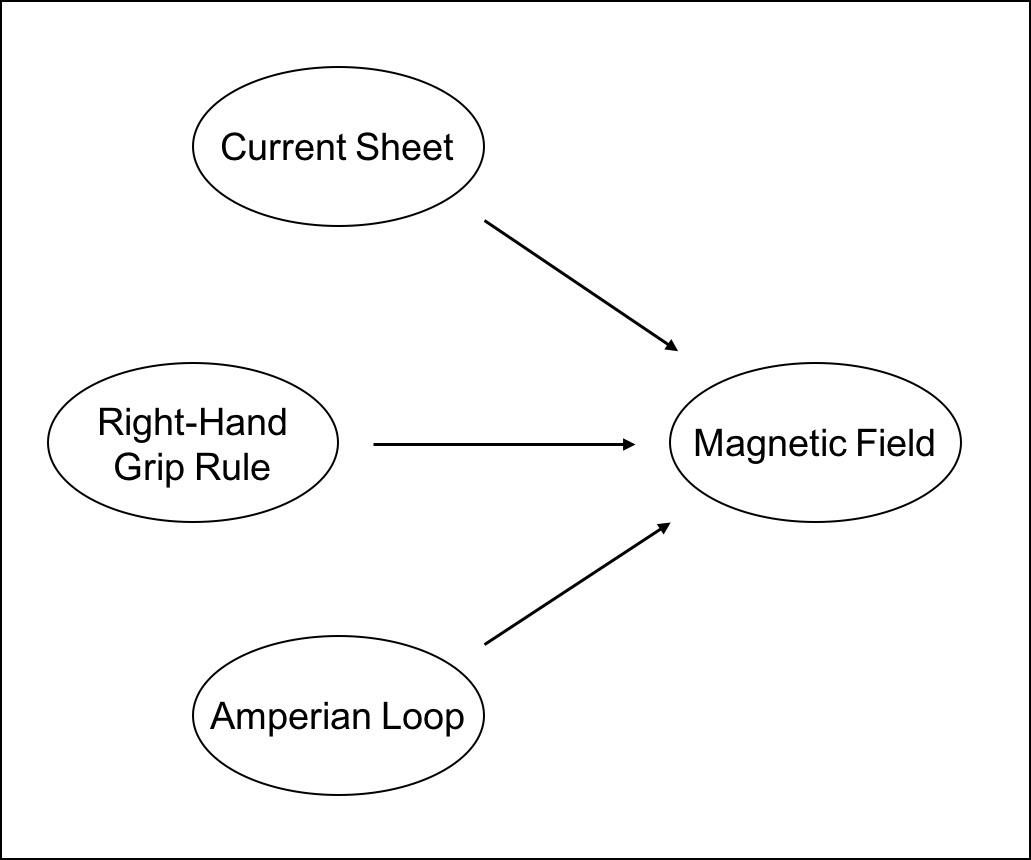}}
\caption{\label{figure9} Three compound representations from Larry's previous episodes combined to give the features (direction and magnitude) of the magnetic field.}
\label{f7}
\end{figure*}

There are three compound representations which give the features (direction and magnitude) of the magnetic field to solve the given problem (Figure 9). First, Larry builds up the compound representation of the current sheet (episodes 1 and 2) that he uses to buildup the direction and the magnitude of the magnetic field. The iconic parts of the current sheet are represented using a parallelogram with a line to represent the current wire, a sheet of paper and some hand gestures. In this compound representation, the parallelogram and the sheet of paper (Table 4) allow Larry to visually represent the current sheet. The inclusion of hand gestures and a line to represent the wire helps to simplify the structure of the current sheet. Then, he does some different things with the right-hand grip rule to buildup magnetic field direction. First, the compound representation of the right-hand grip rule (episodes 1 and 2) is developed using a gesture and a drawing of a line to represent the current-carrying wire with an arrow to represent given current direction. Then, the curled right-hand gesture visualizes the application of the right-hand grip rule. Later, Larry builds up the compound representation of the Amperian loop (episode 3) using gesture and a drawing of a square on the board with an arrow to represent the direction on the loop. In this compound representation, gesturing and drawing (square) on the board (Table 4) allow Larry to visually represent the existence of the loop with respect to the orientation of the current sheet. Then, the inclusion of the direction on the loop helps him figure out the relative orientation with the magnetic field direction.

Table 4: Disciplinary affordances of three representations incorporated to find direction and magnitude of the magnetic field.
\begin{tabularx}{1\textwidth}{|l|X|}
\hline
 Compound Representation & Disciplinary Affordance \\
 \hline
Current sheet & In this compound representation, the sheet of paper and the drawing of the parallelogram on the board generate visual representations of the current sheet. Larry keeps referring back to the diagram on the board as he progresses on this task, and it allows Larry to add different features: current direction (episode 1), current-carrying wire (episode 2), Amperian loop (episode 3) and findings: magnetic field direction (episodes 1 and 2) on to his compound representation. The inclusion of a line helps Larry to visually represent the current-carrying wire in the space of the diagram (on the board) and this allows him to consider multiple current-carrying wires while figuring out the net magnetic field above and below the sheet (episode 2).\\
\hline
Right-hand grip rule & Right-hand rule helps to define the magnetic field. It reveals the connection between current direction and the magnetic field lines in the magnetic field created by a current. In this situation right-hand grip allows Larry to manipulate this phenomenon by hand and to visualize the resulting magnetic field direction. In order to apply the right-hand grip rule, Larry must have a certain current direction. Early in the episode 1, we observe the missing detail of current direction prevents Larry from making a conclusion while reasoning using the diagram on the board. But, the inclusion of the current direction and the line to specify the current wire in episode 2 helps Larry to figure out the magnetic field direction above and below sheet.\\
\hline
Amperian loop & In this compound representation, gesturing and drawing (square) on the board allow Larry to visually represent the existence of the loop with respect to the orientation of the current sheet. We observe these visual representations do not help Larry to simplify the left-hand side of the integral form of the Ampere's law equation. Then, the inclusion of the direction on the loop helps to figure out the relative orientation of the unit length on loop with the magnetic field direction. This leads Larry to find the the magnitude of the magnetic field. \\
 \hline
\end{tabularx}

The compound representation of right-hand grip rule allows Larry to figure out the resulting direction of the magnetic field lines from a current-carrying wire. In order to apply the right-hand grip rule, Larry has to have a specified current-carrying wire with a certain current direction. Larry's approach to first, simplify the current sheet into individual wires and then to focus on a single wire to apply the right-hand grip rule helps to build a valid argument towards finding the net magnetic field direction above and below the sheet. The use of the parallelogram (episode 1) to visually represent the loop in the space of the diagram (on the board) helps Larry to include the arrow to represent the direction on the Amperian loop. This allows Larry to figure out the relative orientation between the magnetic field direction and the unit length on the Amperian loop. Finally, Larry builds a complete argument and advances with the mathematical manipulation to simplify the left-hand side of the integral form of the Ampere's law equation. Then, he completes the oral exam after recording the magnitude of the magnetic field.

The case of Larry exemplifies the coordination between multiple semiotic resources with different disciplinary affordances to build up compound representations to solve complex physics problems. Our analysis of this case illustrates a novel way of thinking about what it means to solve physics problems and, we hope, contributes to the application of social semiotics to the teaching and learning of university physics \cite{Airey2009,chapter5}. This class has a strong focus on problem-solving and sense-making. The instructor involved in this study conducts the oral exam and provides hints to Larry in a certain way that aligns with their practices in the classroom, but we think this interaction does not materially change our argument about how semiotic resources can come together to build representations and arguments to solve problems. Having a single student and an instructor is the limitation of our study, but we note that this is an exemplary case with an existing approach that is worth paying attention to.

\section{Implications \label{sec:implications}}
The process of constructing an effective representation of a problem makes it easier for the problem solver to make appropriate decisions about the solution process. In addition, this process symbolizes the student's work on that particular problem. If the student constructs an effective representation then the student is more likely to progress towards solving the problem \cite{mason_helping_2010}, but if the student constructs an inappropriate representation, then the process is unlikely to make any progress until the student re-represents the problem accurately. This is evident in Larry's case, he could not continue to consider multiple current-carrying wires while using the sheet of paper representation (episode 2), but his decision to switch to the diagram on board helps him to consider multiple wires and leads him to draw a conclusion about the net magnetic field above and below the sheet. 

Further, during episode 3, either gesturing for the loop or recording the loop does not help Larry to figure out the relative orientation between the magnetic field direction and the unit length on the Amperian loop until he adds the direction on the loop. In some cases, students get stuck and cannot identify the nature of the sticking point. We observe in some occasions, Larry gets stuck and his voice level, choice of words and the speed of gestures indicates his confusion. But after adding new features or switching to a different semiotic resource to re-represent the idea or concept Larry's reasoning goes back to normal, and he moves forward to solve the problem. As the case of Larry demonstrates, one of the important problem-solving skills is that of effectively representing and re-representing problems. This skill includes both students and teachers being aware of the nature of the disciplinary affordances of semiotic resources that students bring together to construct representations\cite{fredlund_exploring_2012-1}. 

The research findings presented above suggests that it is important to highlight the complex use of multiple semiotic recourses in student problem solving. An implication is that instructors need to identify the disciplinary affordances of the different semiotic resources in different modalities so that instructors could demonstrate and help students to become better problem solvers. 

\section{Acknowledgments}
This work is supported by National Science Foundation grant 1430967. We are deeply grateful to the members of the K-SUPER group for assistance on this paper and for helpful feedback.

\section{References}
\bibliographystyle{ieeetr.bst}
\bibliography{bibliography.bib}

\bigbreak
\bigbreak
\bigbreak

%\includepdf[pages=-]{Appendix.pdf}
\end{document}